\documentstyle[prl,aps,multicol,epsf]{revtex}

\begin{document}
\draft

\title{Using single quantum states as spin filters to study
spin polarization in ferromagnets}

\author{Mandar M. Deshmukh and D. C. Ralph}
\address{Laboratory of Atomic and Solid State Physics, Cornell
University, Ithaca, NY 14853}
\date{\today}
\maketitle

\begin{abstract}

By measuring electron tunneling between
a ferromagnet and individual
energy levels in an aluminum quantum dot,
we show how spin-resolved quantum states
can be used as filters to determine spin-dependent
tunneling rates.
We also observe
magnetic-field-dependent shifts in the magnet's
electrochemical potential relative to the dot's
energy levels.
The shifts vary between samples and
are generally smaller than expected
from the magnet's spin-polarized
density of states.
We suggest that they are affected by
field-dependent charge
redistribution at the magnetic interface.
\end{abstract}

\pacs{PACS numbers: 73.23.Hk, 75.50.Cc, 75.70.Cn}

\begin{multicols} {2}
\narrowtext
Quantum dots are useful for studying electron
spins, because they allow individual spin-resolved states
to be examined in detail.  Previous experiments have probed spin
physics within several types of quantum dots: semiconductors
\cite{mceuen,tarucha,stewart,ciorga}, nonmagnetic metals
\cite{ralph95,petta},
carbon nanotubes \cite{cobden}, and ferromagnets
\cite{gueron}. Here we use the individual
spin-resolved energy levels
in a quantum dot to investigate
the physics of a bulk magnetic electrode.
The spin polarization in the magnet affects
electron tunneling via the dot levels in two ways.
First, tunneling rates are
different for spin-up and spin-down electrons; we
demonstrate how the
tunneling polarization can be measured by using
quantum-dot states as spin filters \cite{loss}.
Second, as a function of magnetic
field, the electrochemical potential of the magnetic
electrode shifts relative to the energy levels in the dot.
Previously, tunneling polarizations \cite{tm} and electrochemical
shifts \cite{ono} have been measured by other techniques in
larger devices having continuous densities of electronic states.
By probing at the level of single quantum states, we are able to
compare both effects in one device.  We also achieve
more precise measurements of the electrochemical shifts
which demonstrate that they are not determined purely by the
bulk properties of the magnet,
as has been assumed previously \cite{ono,macdonald}.

Our quantum dot is an Al particle, 5-10~nm in diameter,
connected by Al$_2$O$_3$ tunnel junctions to an Al electrode on
one side and a cobalt or nickel electrode on the other
(Fig.~1, inset).
We use an Al particle to
minimize spin-orbit coupling, so that electronic states within the
particle are to a good approximation purely spin-up or spin-down
\cite{ralph95,petta}. Device fabrication is done following the
recipe in \cite{ralph95}, except that in the final step we deposit
80~nm of magnetic Co or Ni at a pressure of $\sim
2 \times 10^{-7}$ torr to form the second electrode. We conduct
tunneling measurements in a dilution refrigerator, using filtered
electrical lines that provide an electronic base temperature of
approximately 40~mK. Beyond a threshold voltage determined by the
charging energy, electron tunneling via individual quantum states
in the Al particle produces discrete steps in the $I$-$V$ curve
\cite{ralph95} or equivalently peaks in $dI/dV$ vs.\ $V$ (Fig.\
1). The sign of bias refers to the sign applied to the Al
electrode.  Figure 2 shows how the energy levels in the particle
undergo Zeeman spin-splitting as a function of magnetic field
  ($B$, applied in the plane of the nitride membrane)
\cite{ralph95}. The Co-lead sample also exhibits nonlinearities
for $B <$ 0.3 T, possibly associated with magnetic-domain
rotation.
\vspace{-0.25 cm}
\begin{figure}
\begin{center}
\leavevmode \epsfxsize=7.5 cm \epsfbox{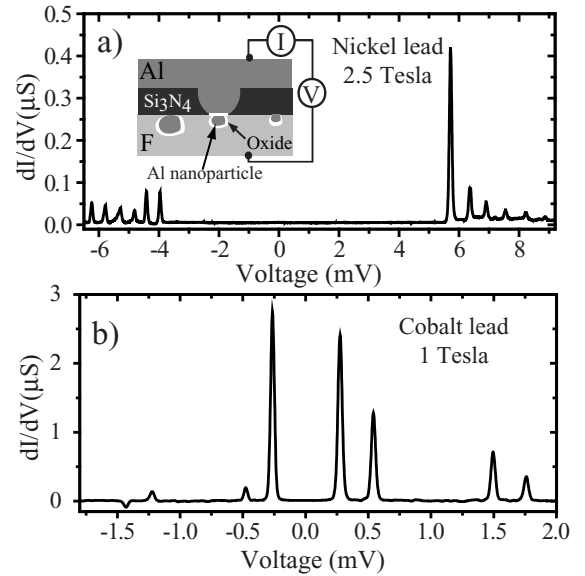}
\end{center}
\vspace{-0.25 cm} \caption{ \label{figure1} (inset)
Cross-sectional device schematic. (a)  Differential conductance
vs.\ $V$ for device Ni\#1 with one Ni electrode, and (b) for
device Co\#1 with one Co electrode. Magnetic fields are applied to
cause Zeeman splitting of the spin-up and spin-down resonances.}
\end{figure}

Before we turn to our main results, we note some experimental
details.  In order to convert the measured voltages of the
resonances to energy, one must correct for the capacitive division
of $V$ across the two tunnel junctions. For a tunneling transition
across the nonmagnetic (N) junction, this is accomplished by
multiplying $V$ by $eC_F/(C_N+C_F)$ and for the ferromagnetic (F)
junction by $eC_N/(C_N+C_F)$, where $C_N$ and $C_F$ are the two
junction capacitances. The capacitance ratio can be determined by
comparing the voltage for tunneling through the same state at
positive and negative $V$ \cite{ralph95}.  We must also understand
whether a resonance corresponds to a threshold for an electron
tunneling on or off the particle, and across which tunnel
junction. The transitions which correspond to tunneling between
the particle and the Al electrode can be identified by the
presence of a shift in their $V$ positions as the Al electrode is
driven from superconducting to normal by a magnetic field, and by
the effect of the superconducting density of states (DOS) on the
resonance shape \cite{ralph95}. The sign of $V$ then determines
whether an electron is tunneling on or off the particle. For the
sample (Ni\#1) shown in Figs.\ 1(a) and 2(a,b), the transitions at
positive $V$ correspond to tunneling first from the dot to the Al
electrode, with $eC_F/(C_N+C_F)$ = (0.42 $\pm$ 0.02)e.  For the
sample (Co\#1) in Figs.\ 1(b) and 2(c), at positive $V$ electrons
are initially tunneling from the Co electrode to the particle, and
$eC_N/(C_N+C_F)$ = (0.44 $\pm$ 0.01)e.
\vspace{-0.25 cm}
\begin{figure}[t]
\begin{center}
\leavevmode \epsfxsize=7.5 cm \epsfbox{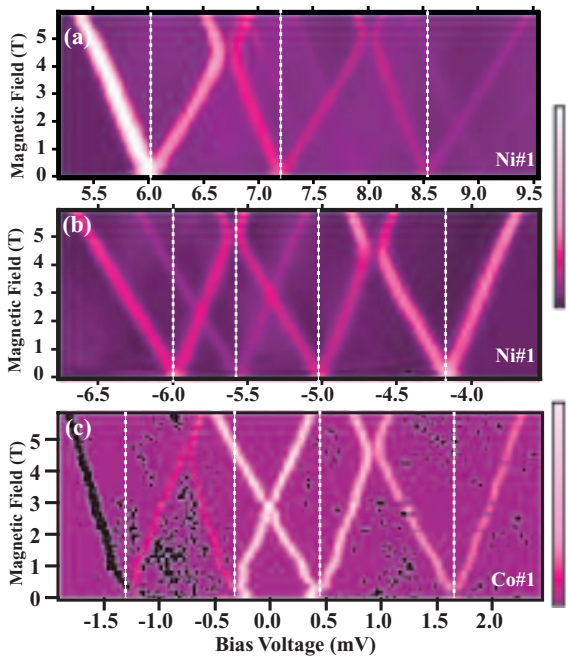}
\end{center}
\vspace{-0.25 cm} \caption{ \label{figure2} $dI/dV$ vs.\ voltage
and magnetic field for (a,b) device Ni\#1 and (c) Co\#1. The
scales extend from 0 to (a,b) 0.2 $\mu$S and (c) 2 $\mu$S. White
indicates $dI/dV$ values beyond the scale maximum, and in (c)
black indicates negative values. }
\end{figure}

We will now analyze how the currents carried by individual states
allow measurements of spin-dependent tunneling rates. The
resistances of our tunnel junctions are sufficiently large (at
least 1 M$\Omega \gg h/e^2$) that transport can be modeled by
sequential tunneling \cite{vondelft,bonet}.  The analysis takes a
particularly simple form when the offset charge \cite{vondelft}
has a value that permits tunneling at a low value of $V$ so that
only a single orbital state on the quantum dot contributes to
current flow near the tunneling threshold \cite{bonet,deshmukh2}.
This is the case for sample Co\#1; the thresholds for more
complicated non-equilibrium tunneling processes, involving the
lowest-energy even-electron excited state \cite{loss}, are $V<$
-5.8 mV or $V>$ 4.2 mV at $B\!=\!0$ in this sample. In general the
simple equilibrium tunneling regime can be achieved for any
nanoparticle device made with a gate electrode so that the offset
charge can be adjusted \cite{deshmukh2}. In Fig.\ 3(a) we show the
$I$-$V$ curve for sample Co\#1 with $B$ = 1 T to Zeeman-split the
resonances. The first step in current for either sign of $V$
corresponds in this sample to an electron tunneling through only a
spin-up (majority-spin) state. The sequential tunneling theory
\cite{bonet} predicts that these two currents should have
identical magnitudes,
\begin{equation}
I_{1+} = |I_{1-}| = e \gamma_{\uparrow} \gamma_N /
(\gamma_{\uparrow}+ \gamma_N),
\end{equation}
where $\gamma_{\uparrow}$ is the bare tunneling rate between the
magnet and the spin-up state, and $\gamma_N$ is the tunneling
rate to the Al electrode. The fact that the steps do
have the same magnitude confirms that electrons are
tunneling via just one state. When $|V|$ is increased
to permit tunneling through either
the spin-up or spin-down state, the
predicted values for the total current, using the methods in
\cite{vondelft,bonet}, are for positive and negative $V$,
\begin{equation}
I_{2+}=\frac{e\gamma_N (\gamma_{\uparrow} + \gamma_{\downarrow})}
{\gamma_N + \gamma_{\uparrow} + \gamma_{\downarrow}},
\end{equation}
\begin{equation}
|I_{2-}|=\frac{2e\gamma_N}{1 + \gamma_N/\gamma_{\uparrow} +
\gamma_N/\gamma_{\downarrow}}.
\end{equation}
We have made use of time-reversal symmetry which requires that the
tunneling rates from the nonmagnetic electrode to both
Zeeman-split states should be the same.
This has been verified in a previous
experiment \cite{deshmukh2}.
We have also neglected spin relaxation
based on experimental limits of relaxation rates slower than
5$\times$10$^7$ s$^{-1}$ in Al particles with Al
electrodes \cite{deshmukh2}, much slower than the
tunneling rates.
Equations (1)-(3) can be inverted to determine $\gamma_N$,
$\gamma_{\uparrow}$,
and $\gamma_{\downarrow}$ from
$I_{1+}$, $I_{2+}$, and $I_{2-}$ (Fig. 3(c)). The resulting
tunneling polarization, $(\gamma_{\uparrow}-\gamma_{\downarrow}) /
(\gamma_{\uparrow}+\gamma_{\downarrow})$, is positive (Fig.~3(d)),
meaning that the tunneling rate for spin-up (majority) electrons
in the ferromagnet is faster than for spin-down. This
sign agrees with
results for tunneling
from ferromagnets through Al$_2$O$_3$ into thin-film
superconducting Al (FIS devices) \cite{tm,mp},
although the sign is opposite to the polarization of the DOS
at the Fermi level within band structure calculations
\cite{papa}. This is understood to be due to much larger
tunneling matrix elements for predominantly $sp$-band
majority-spin electrons compared to predominantly $d$-band
minority electrons, so that the matrix elements dominate over the
DOS effect in determining the relative tunneling
rates \cite{stearns,tp}. The magnitude of the tunneling
polarization that we measure (8-12\%) is considerably less than
the values 35-42\%
found for Co using FIS devices \cite{tm,mp},
and we observe some field dependence
not seen in larger samples (Fig.\ 3(d)).
Both effects may indicate imperfections in our tunnel
barriers; they
have not undergone
the process of optimization which achieved large polarizations in
larger-area devices \cite{seve}.  The presence of any oxidation at the
magnetic interface can reduce the tunneling polarization \cite{seve}.
Our barriers are also
very thin (with resistance-area products
less than 200 $\Omega$-$\mu$m$^2$ compared to 10$^7$-10$^{10}$
$\Omega$-$\mu$m$^2$ in most prior experiments \cite{tm,monsma}),
which might reduce the polarization by increasing the
relative tunneling rate of $d$-states \cite{oleinik}.  (However,
recent work on optimized large-area F/Al$_2$O$_3$/F tunnel junctions with
$RA \sim$ 100 $\Omega$-$\mu$m$^2$ does
not show reduced polarization \cite{monsma}.)
We have
considered whether the magnetic electrode might enhance
spin relaxation within the particle  so that it should not be neglected.
This cannot explain the full reduction in our polarization; treating the
relaxation rate as a free parameter, the maximum polarization
consistent with the current steps in Fig.\ 3(a) is 21\%.
\vspace{-0.3cm}
\begin{figure}
\begin{center}
\leavevmode \epsfxsize=7.0cm \epsfbox{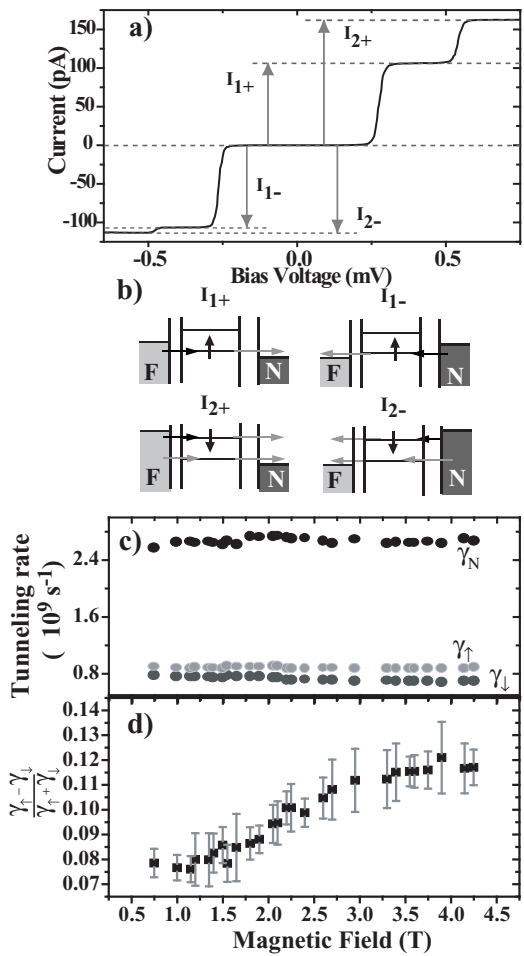}
\end{center}
\vspace{-0.35 cm} \caption{ \label{figure3} (a) Current vs.\
voltage curve for device Co\#1 for $B$ = 1 T, showing the range of
$V$ where tunneling occurs via one pair of Zeeman-split energy
levels. (b)  Energy-level diagrams at each current step. Black
horizontal arrows show the threshold tunneling transition. Gray
arrows depict other transitions which contribute to the current.
(c) The tunneling rates $\gamma_{\uparrow}$,
$\gamma_{\downarrow}$, and $\gamma_N$ determined as described in
the text. (d)  Tunneling polarization for device Co\#1.}
\end{figure}

In Figs.\ 1(b) and 2(c) at negative $V$, some higher-energy
spin-down resonances produce $dI/dV\!<\!0$, meaning that they
decrease the total current. This is a consequence of the slower
rate of tunneling for minority-spin electrons; an electron in the
spin-down state blocks current flow through the spin-up channel
until the electron tunnels slowly to the F electrode.  By
incorporating additional states into the sequential-tunneling
model, we find a tunneling polarization of $15 \pm 6\%$ for the
second Zeeman pair in sample Co\#1.

Let us now consider the $V$ positions of the tunneling resonances
as a function of $B$. The magnitude of the Zeeman splitting is
similar to previous measurements in all-Al devices \cite{ralph95}.
After converting from $V$ to energy as described above, we
determine the g-factor according to $\Delta E_{Zeeman} = g \mu_B
B$.  For the levels in Ni\#1,  $g$ is between 1.83 $\pm$ 0.05 and
1.90 $\pm$ 0.07, in Co\#1 between 1.98 $\pm$ 0.07 and 2.05 $\pm$
0.06, and in the other devices discussed in this paper, 1.9 $\le g
\le 2.0$. However, the data in Fig.\ 2 differ from studies with
non-magnetic electrodes \cite{ralph95,deshmukh2} in that the
slopes of the spin-up and spin-down Zeeman shifts are not
symmetric about 0; in Figs.\ 2(a,b) the midpoints of the
Zeeman-split states tend to higher $|V|$ as a function of $B$ and
in Fig.\ 2(c) they tend to lower values of $|V|$. This effect is
expected in single-electron transistors (SETs) made with magnetic
components, as a result of a field-dependent change in a magnet's
electrochemical potential \cite{ono,macdonald}. A related shift
has been observed in micron-scale Ni/Co/Ni, Co/Ni/Co, and Al/Co/Al
SETs \cite{ono}.   When a magnetic field is applied to any bulk
metal, it will flip some electron spins to align with $B$. Because
a ferromagnet has different densities of states at the Fermi level
for majority- and minority-spin states, the electrochemical
potential must shift with B to accommodate the flipped spins. The
magnitude of the shift will also be enhanced by exchange
interactions in the magnet \cite{macdonald}. We will parameterize
the shift by the variable $S$, such that for an isolated magnetic
sample $\Delta E_F(B) = S \mu_B B$. When a magnet is incorporated
as one electrode in an otherwise nonmagnetic SET, the experimental
consequences of this shift are equivalent to a change in the
energy of all the states in the nanoparticle by the amount $dE/dB
= -\mu_B S C_F/(C_N + C_F)$ \cite{ono}. This analysis assumes that
the magnetic field does not induce any rearrangements of charge
density (see below).

Within each sample, the average slopes of the different
Zeeman-split pairs correspond to the same value of $S$
within measurement uncertainty.  For the data in Figs.~2(a,b), the
average slopes are
(2.6 $\pm$ 0.2) $\times$ 10$^{-2}$ mV/Tesla for positive $V$
and
(1.85 $\pm$ 0.2) $\times$ 10$^{-2}$ mV/Tesla for negative $V$,
giving in both cases $S$ = $0.45~\pm~0.04$. For
non-interacting electrons with different majority and
minority densities of states per unit energy
at the Fermi level, $\rho_{\uparrow}$
and $\rho_{\downarrow}$, the DOS polarization would
give a shift $S=-(1/2) g
[(\rho_{\uparrow}-\rho_{\downarrow})/(\rho_{\uparrow}+
\rho_{\downarrow})]$, where $g$ is the $g$-factor
\cite{ono,macdonald}. Therefore a positive sign for $S$
corresponds to a greater density of
minority-spin states at the Fermi level, in agreement with
band structure calculations
for Ni and Co \cite{papa}.  However, the magnitude of the measured
shift is surprisingly small. Band-structure calculations for Ni
give $\rho_{\downarrow}/\rho_{\uparrow}$ = 8.5 \cite{papa}, so that
one would expect $S >$ 0.79.  We write this as a lower limit,
because exchange interactions should increase
$S$ relative to predictions for
non-interacting electrons \cite{macdonald}.
For two other devices with
a Ni electrode, made by the same procedure, we find
even more striking discrepancies: $S$ =
0.15 $\pm$ 0.1 and 0.2 $\pm$ 0.1.  For 3 samples with a Co
electrode, for which band structure calculations suggest that $S
>$ 0.59 \cite{papa}, we observe $S$ = 0.1 $\pm$ 0.1,
0.37 $\pm$ 0.05 (for Co\#1), and 0.7 $\pm$ 0.1.
Even though the electrochemical
shift is expected to be a bulk property of the magnet
\cite{ono,macdonald},
we find significant
sample-to-sample variations for the shift in the magnet's
electrochemical potential relative to the Al
particle.

We propose that the explanation of these discrepancies is that a
magnetic field may produce rearrangements in the charge distribution
inside a magnetic tunnel junction that will
shift the energy levels of the particle as a function of $B$,
with different strengths in each device.
The predominantly $d$-band character of the minority electrons in
Co or Ni will cause their wavefunctions to decay over a shorter distance as
they penetrate into the tunnel barrier than for the predominantly
$sp$-band majority electrons \cite{oleinik}.
Therefore, as an applied magnetic field transfers
electrons from minority to majority states, some charge density at
the surface of the magnet should shift slightly toward the barrier
region \cite{zhang}.  More complicated spin-dependent
interface states
could act similarly.
The sign of the effect should cause the measured values of $S$ to decrease
for Ni and Co electrodes, and the
magnitude will be
given by the work that the moving charge does on an
electron in the particle.
Making the
approximation that the spin-dependent densities at the
magnet's surface are similar to the bulk, the charge density
per unit area which changes spin at the last monolayer of
the magnet is $\sigma \approx e a \rho_{\uparrow}
\rho_{\downarrow} g \mu_B B/(\rho_{\uparrow}+
\rho_{\downarrow})$, where $a$ is the lattice constant.
If the average position for charges in the minority and
majority states differs by
$\Delta x$ at the surface layer, then this charge movement
should change the measured
electrochemical shift by
\begin{eqnarray}
\Delta S & \approx &  - \frac{e^2}{\epsilon_0} g a (\Delta x)
\frac{\rho_{\uparrow}
\rho_{\downarrow}}{\rho_{\uparrow} + \rho_{\downarrow}} \\
& \approx & - 12 \Delta x /\AA
\end{eqnarray}
for either a Co or Ni electrode.
Therefore even in micron-scale devices \cite{ono},
$\Delta x$ as small as 0.01 $\AA$ may decrease $S$
by 10\%, and foil attempts to measure the
DOS polarization.  In our devices,
which have possibly non-uniform tunnel barriers,
variations in $\Delta x$ by less than 0.05
$\AA$ can explain the
sample-to-sample differences.

In summary, we have studied tunneling between a bulk
ferromagnet and the spin-resolved energy levels in a quantum dot.
The energy levels can be used as spin filters, permitting
a measurement of the
different tunneling rates from the magnet for
spin-up and spin-down electrons.
As a function of $B$, the electrochemical potential in the
magnet shifts relative to the energy levels in the quantum dot.
In addition to the shift that is expected due to the magnet's bulk
density-of-states polarization \cite{ono},
we suggest that there is an important contribution from $B$-dependent
redistributions of charge at the magnetic interface.

We thank R. A. Buhrman, A. Champagne, S. Gu\'eron, A. H.
MacDonald, D. J. Monsma, and E. B. Myers for discussions.
Funding was provided by ARO
(DAAD19-01-1-0541), the Packard Foundation, and NSF
(DMR-0071631 and use of the National Nanofabrication Users
Network).

%
%
%

\end{multicols}

\end{document}